# The Oxygen Reduction Pathway for Spinel Metal Oxides in Alkaline Media: An Experimentally Supported *Ab Initio* Study

Colin R. Bundschu[1*], Mahdi Ahmadi[2], Juan F. Méndez-Valderrama[3], Yao Yang[2,4], Héctor D. Abruña[2*], Tomás A. Arias[3*]

1) Department of Applied and Engineering Physics, Cornell University, Ithaca, NY 14853, USA.
2) Department of Chemistry and Chemical Biology, Cornell University, Ithaca, New York 14850, United States
3) Department of Physics, Cornell University, Ithaca, NY 14853, USA
4) Present Address: Department of Chemistry, Miller Institute, University of California, Berkeley, Berkeley 94720, CA.

*Corresponding authors. e-mail addresses: crb273@cornell.edu, hda1@cornell.edu, taa2@cornell.edu

## Abstract

Precious-metal-free spinel oxide electrocatalysts are promising candidates for catalyzing the oxygen reduction reaction (ORR) in alkaline fuel cells. In this theory-driven study, we use joint density-functional theory in tandem with supporting electrochemical measurements to identify a novel theoretical pathway for the ORR on cubic $Co_3O_4$ nanoparticle electrocatalysts. This pathway aligns more closely with experimental results than previous models. The new pathway employs the cracked adsorbates *(OH)(O) and *(OH)(OH), which, through hydrogen bonding, induce spectator surface *H. This results in an onset potential closely matching experimental values, in stark contrast to the traditional ORR pathway, which keeps adsorbates intact and overestimates the onset potential by 0.7 V. Finally, we introduce electrochemical strain spectroscopy (ESS), a groundbreaking strain analysis technique. ESS combines ab initio calculations with experimental measurements to validate proposed reaction pathways and pinpoint rate-limiting steps.

## Introduction

The high cost of platinum alloys and the sluggish kinetics of the oxygen reduction reaction (ORR) at the cathode present major hurdles to the commercialization of both acidic and alkaline fuel cells.[1–6] Great attention has been devoted to finding non-precious metal electrocatalysts which could resolve these issues. In this search, metal-nitrogen doped carbons (M-N/C),[7] spinel oxides (Co, Mn, Fe)[8–10] and perovskites[11] are of considerable interest in alkaline fuel cells due to their faster ORR kinetics and high stability in alkaline environments.[12]

In the case of spinel oxides, numerous strategies have been attempted to optimize the reactivity and stability of catalysts for the ORR. Xiong *et al.* found that in a Mn-doped cobalt ferrite, a synergistic effect between Co and Mn enhances the ORR reactivity while Fe stabilizes the catalyst by preserving the spinel structure.[8] Wei *et al.* explored the effect of Mn valency on activity and found that the ORR activity of $MnCo_2O_4$ produces a volcano shape as a function of the Mn valence state, peaking at a valence of +3.[13] Li *et al.* showed that the crystalline structure of spinel $CoMn_2O_4$ affects its reactivity, with higher activity in cubic $CoMn_2O_4$ than tetragonal $CoMn_2O_4$.[14] Of the numerous parameters, the shape and structure of nanoparticle (NP) facets, *in particular,* strongly affect ORR performance.[15–17] For example, it was found that ellipsoidal $Mn_3O_4$ NPs achieve higher ORR activity than spherical and cubic NPs,[9] while Gao *et al.* showed that $Co_3O_4$ (111) presents a lower activation barrier for $O_2$ desorption as compared to $Co_3O_4$ (100) for the oxygen evolution reaction (OER).[16]

In contrast to these numerous experimental studies focused on improving ORR metal-oxide performance, attempts to elucidate the measured activity using *ab initio* calculations have proved consistently challenging.[18,19] Traditionally, the metal oxide alkaline ORR pathway has been believed to be *OH→*OO→*OOH→*O→*OH. [20] However, when tested using the simple (100) facet of $Co_3O_4$, this traditional pathway produces energies in disagreement with experiment. Specifically, as we show below, joint density-functional theory calculations at the experimental onset potential ($V_{on}$) for $Co_3O_4$ (100) that we measure below, and which others have observed in cyclic voltammetry experiments,[23] indicates that the traditional pathway leads to unphysical reaction barriers exceeding 0.7 eV, with the most favorable path still requiring kinetically challenging hopping of intermediates between surface sites.

Given the large discrepancies between the computed and experimental energies for the simple $Co_3O_4$ (100) test case, the traditional pathway very likely misses key reaction intermediates. Using the traditionally assumed pathway as a starting point, and drawing inspiration from known platinum group metal (PGM) catalyst reactions, we perform new JDFT calculations and propose that a more accurate alkaline ORR pathway on the $Co_3O_4$ (100) surface is as presented in Figure 1. Not only does this pathway first begin to present no significant barrier at the experimentally observed onset potential**,** it also appears to be kinetically feasible with all intermediates remaining on a single site. Finally, with this new pathway in hand, below we further demonstrate that the ORR activity of cubic $Co_3O_4$ nanoparticles can be tuned significantly by controlling the nanoparticle strain.

**\*H + \*OH + 2H$_2$O + O$_2$ + 4e$^-$** → * + *(OH)(O) + 2H$_2$O + OH$^-$ + 3e$^-$

\* + *(OH)(O) + 2H$_2$O + OH$^-$ + 3e$^-$ → * + *(OH)(OH) + H$_2$O + 2OH$^-$ + 2e$^-$

\* + *(OH)(OH) + H$_2$O + 2OH$^-$ + 2e$^-$ → *H + *(OH)(OH) + 3OH$^-$ + e$^-$

\*H + *(OH)(OH) + 3OH$^-$ + e$^-$ → **\*H + \*OH + 4OH$^-$**, (1)

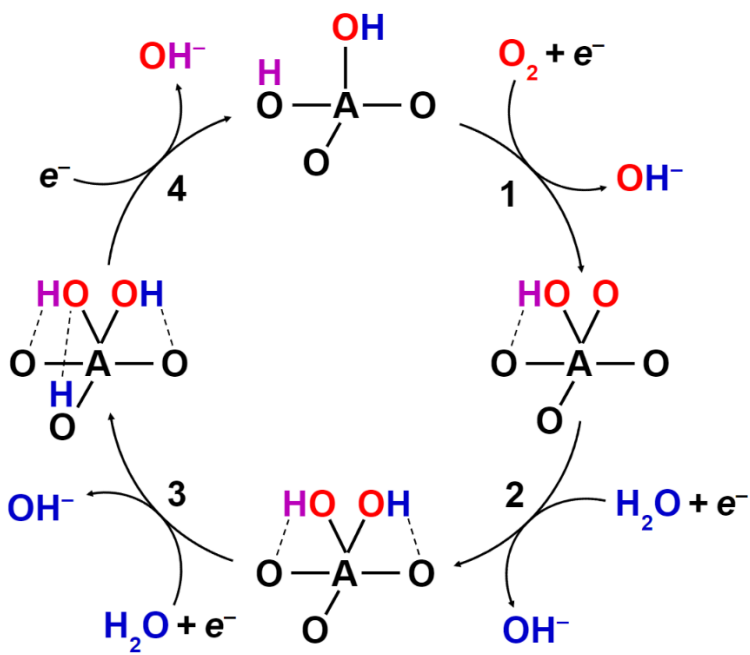

*Figure 1. Schematic of proposed, JDFT-identified ORR pathway with colors differentiating adsorbed O (red), adsorbed H (blue), active spectator H (purple), and the $Co^{2+}$ A-site with participating adjacent surface O sites (black). In transition 4, the blue, surface adsorbed H is equivalent by symmetry to the purple spectator H in the subsequent state, so that the reaction can proceed without need for the reactant to change adsorption site.*

## Results

### Cubic $Co_3O_4$ Nanoparticle Characterization

To test our *ab initio* predictions for the alkaline ORR pathway, we synthesized and characterized cubic (100) $Co_3O_4$ nanoparticles (NPs). Figures 2 (a) and (b) show a TEM image and NPs size histogram of the NPs, respectively, with an average size of 53 ± 16 nm. Figure 2(c) shows the XRD pattern for the NPs alongside the reference single-phase spinel $Co_3O_4$ (PDF 01-076-1802). Figure 2 (d) shows an atomic resolution bright field STEM image of the NPs and (d) shows the corresponding Fourier transform. The FFT image can be indexed as the [110] zone axis of spinel $Co_3O_4$. Figure 2(f) shows the ORR polarization curve for the NPs acquired at a scan rate of 5 mVs$^{-1}$ in 1 M KOH at 1,600 rpm. At potentials higher than ~0.8 V vs. RHE, the current is kinetically controlled, while at lower potentials (< 0.7 V vs. RHE), a diffusion-limited current density of -3.7 mA/cm$^2$ was reached at 1,600 rpm, which matches well with the expected current density of -3.8 mA/cm$^2$ for a 4e$^-$ ORR process at 1,600 rpm in 1 M KOH.[28] Additionally, our onset potential of 0.89 V vs RHE for the cubic surface is in excellent agreement with Yang *et al.*,[23] determined in both cases as the potential yielding 5% of the maximum current.

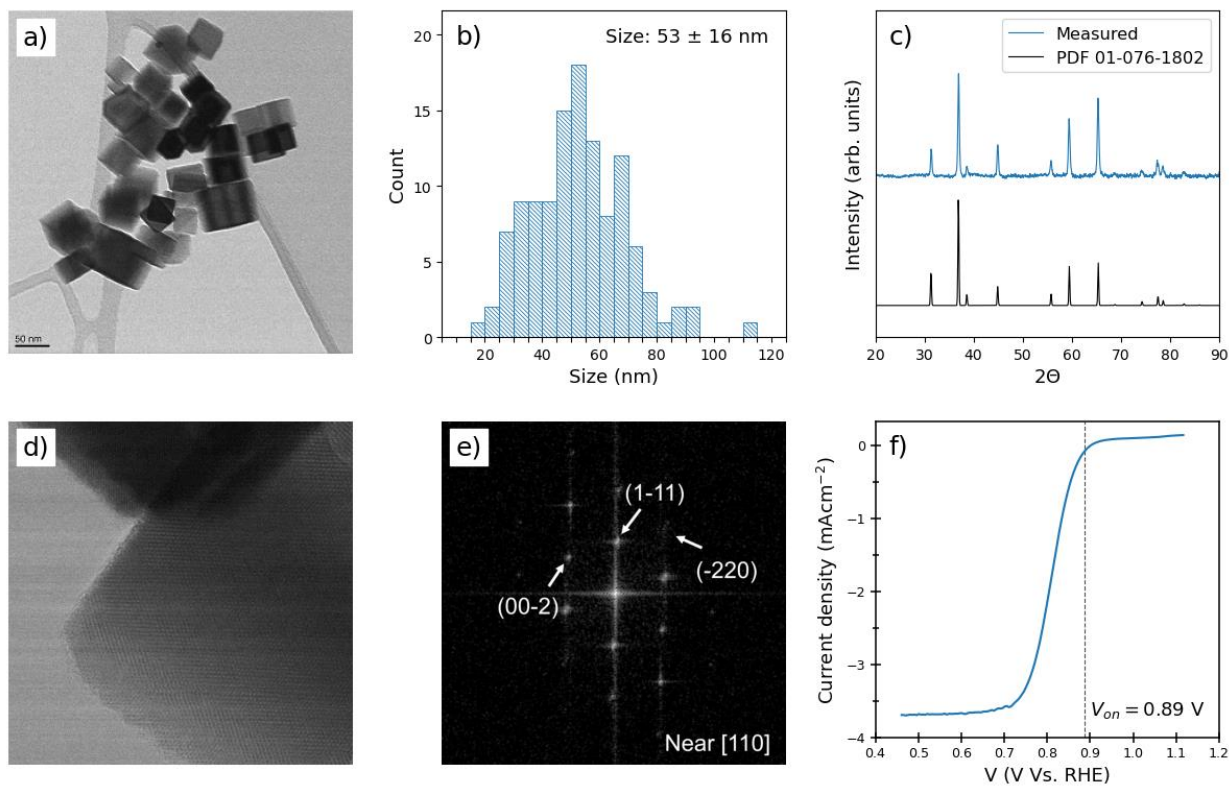

*Figure 2. Characterization of Cubic $Co_3O_4$ Nanoparticles (NPs). (a) Representative TEM image. (b) Particle size distribution histogram. (c) Measured XRD patterns alongside reference single-phase spinel $Co_3O_4$ (PDF 01-076-1802). (d) High resolution BF-STEM image and (e) corresponding Fourier transform. (f) ORR polarization curve in $O_2$-saturated 1 M KOH at 1600 rpm.*

## Identifying the Reaction Pathway

To compute the energies of reaction intermediates on $Co_3O_4$ (100), we employ the most energetically favorable surface termination at the experimental conditions of 298K and atmospheric oxygen partial pressure as determined by Zasada *et al*.[24] The resulting surface exhibits two possible surface Co attachment sites, the protruding tetrahedral (Tet) $Co^{2+}$ and the surface octahedral (Oct) $Co^{3+}$ sites. (See Figure 3a, which presents our *ab* initio results for the solvated, relaxed surface without adsorbates.)

With the surface state established, we first consider the traditional 4e$^-$ pathway for alkaline ORR on metal oxide surfaces,[20]

**\*OH + 2H$_2$O + O$_2$ + 4e$^-$ →**\*OO + 2H$_2$O + OH$^-$ + 3e$^-$

\*OO + 2H$_2$O + OH$^-$ + 3e$^-$ →\*OOH + H$_2$O + 2OH$^-$ + 2e$^-$

\*OOH + H$_2$O + 2OH$^-$ + 2e$^-$ →\*O+ H$_2$O + 3OH$^-$ + e$^-$

\*O+ H$_2$O + 3OH$^-$ + e$^-$ →**\*OH + 4OH$^-$**, (2)

with the start and end states indicated in boldface. In our treatment of this pathway, we compute the energy of each possible intermediate state at both the $Co^{2+}$ and $Co^{3+}$ sites. We also consider possible desorption and readsorption of spectator *H and *OH species during the reaction, finding that the spectator surface H tends to be stabilized by the presence of adsorbed intermediates. For example, we find that whereas, on the clean surface, spectator H's adsorbed to surface oxygen sites desorb at voltages above 0.4 V vs. RHE, the same spectator is stabilized at voltages up to 0.9 V vs. RHE by the presence of the first intermediate of the traditional pathway (Figure 3b).

The top two rows of Figure 3c display our predicted *ab initio* configurations for the traditional pathway using the lowest energy spectator configuration for each reaction intermediate, organized from left to right by reaction step. Each row represents a different attachment site: tetrahedral $Co^{2+}$ (first row) and octahedral $Co^{3+}$ (second row). Choosing the most energetically favorable spectator configuration for each stage of the reaction, at the experimental onset potential ($V_{on}$) of 0.89 V vs. RHE,[23] we find a lower bound to the reaction barrier of 0.71 eV, coming from the *OH → *OO transition (Figure 4a, upper, red pathway). Furthermore, this lowest energy barrier for the traditional pathway requires the adsorbate to jump from the $Co^{2+}$ to the $Co^{3+}$ site. Even if the kinetics associated with this configurational change are feasible, a barrier of this height indicates that the reaction must occur along a different pathway.

To find a more realistic reaction pathway, we have considered a richer set of adsorbate configurations on the surface as shown in rows 3-5 of Figure 3c. Inspired by the split configurations often found for PGM pathways, row three contains cracked configurations where the reaction intermediate remains bonded to a single tetrahedral $Co^{2+}$ site, but with its atoms distributed into different molecular fragments. (We found cracked configurations on the $Co^{3+}$ sites to be unstable and to spontaneously recombine into a single unit.) Moreover, row four contains configurations that involve different atoms from the adsorbate bridging a tetrahedral and an octahedral site. Finally, row five contains configurations where a single atom from the adsorbate sits between the tetrahedral and octahedral sites. In some of these configurations, the stabilized active spectator species H shifts away from its surface site and toward the nearest adsorbed oxygen, as in the configuration labelled * + *(OH)(O) (row three, column two). We denote this configuration as * + *(OH)(O), rather than *H + *(O)(O), to highlight the fact that the hydrogen sits closer to the adsorbed oxygen than to its "home" surface site. The kinetics of this rearrangement more closely resemble the efficient proton transport of the Grotthuss mechanism[25] than they do the less kinetically accessible full hop between sites which would have to occur for the traditionally assumed pathway.

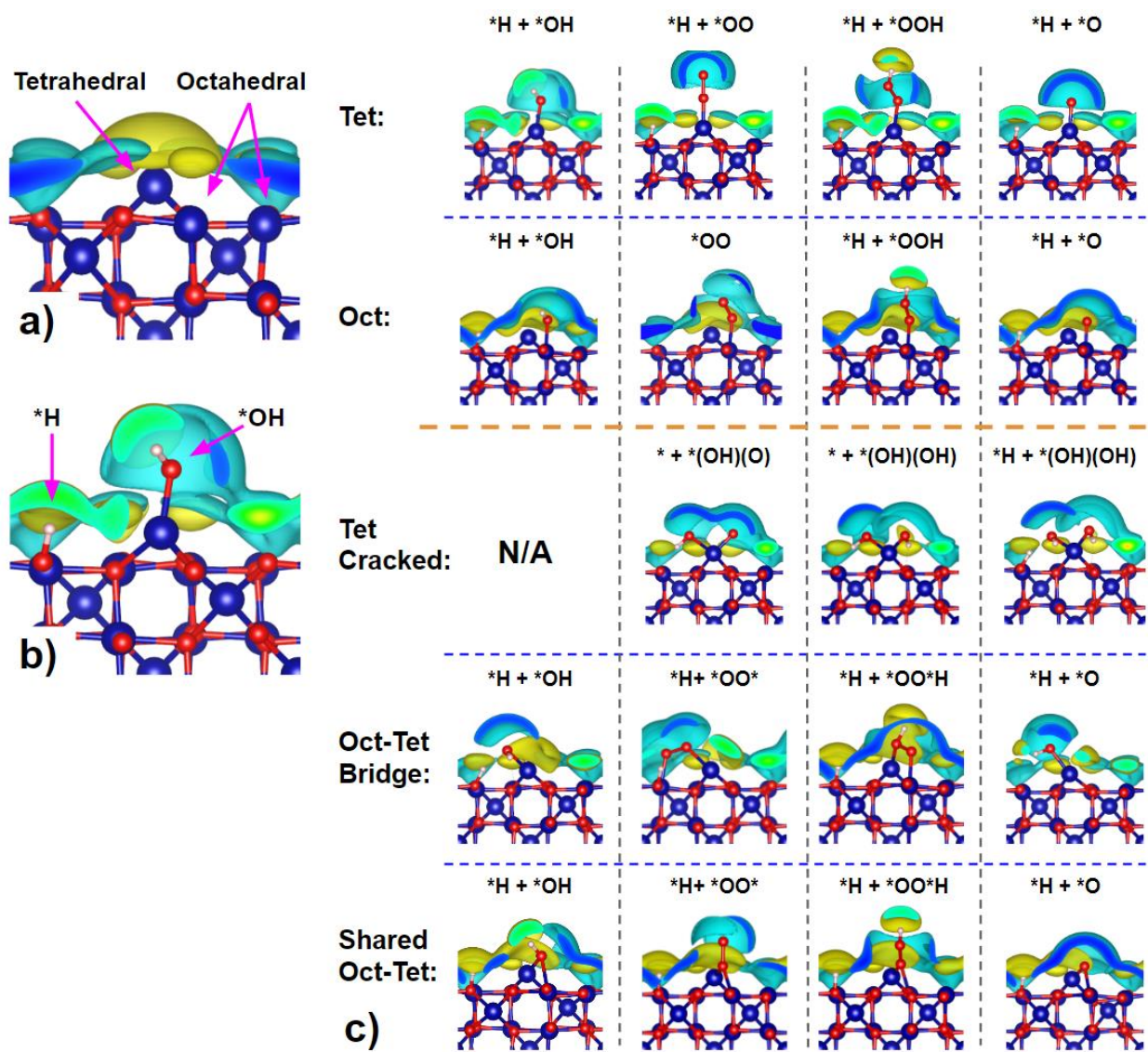

*Figure 3. Solvated JDFT calculations of the (100) surface with a) no adsorbates, b) a spectator *H and an adsorbed *OH, and c) adsorbates organized by reaction step (columns) and adsorbate configuration (rows): Co atoms (blue), O atoms (red), H atoms (white), and positive and negative solvent-charge isosurfaces (blue/cyan and yellow surfaces, respectively). As expected, positive solvent charges (blue/cyan isosurfaces) surround the surface and adsorbed O atoms, whereas negative solvent charges (yellow) surround surface Co and adsorbed H atoms. (Visualizations of solvated surface generated using VESTA.[26])*

At potentials near $V_{on}$, we find the pathway on the $Co^{2+}$ site

$$*H + *OH \;\rightarrow\; * + *(OH)(O) \;\rightarrow\; * + *(OH)(OH) \;\rightarrow\; *H + *(OH)(OH) \;\rightarrow\; *H + *(OH), \quad (3)$$

which starts and ends with the "Tet" *H + *OH configuration and has its intermediates appearing along the row of Figure 3 labeled "Tet Cracked", to be the most energetically favorable and to thereby represent the minimum energy pathway (MEP) for alkaline ORR on the $Co_3O_4$ (100) surface.

Figure 4(a) displays the free energies of the MEP configurations (black) alongside the energies of the traditional pathway (red) at our experimentally observed onset potential $V_{on}$. In correspondence with experiment, the MEP, unlike the traditional pathway, presents no apparent barriers at $V_{on}$. Moreover, at the experimental onset potential, the MEP exhibits transition steps with very small energy drops, consistent with the reaction being near onset. Additionally, this pathway gives insight into a potentially key rate-limiting mechanism for the alkaline ORR on this surface. Specifically, $V_{on}$ marks the point where it becomes energetically favorable for the adsorbed *OH to induce the spectator *H, which is necessary to eliminate barriers along the MEP. This suggests that the reaction onset is not only limited by the transition between any two steps in the reaction pathway, but also by the ability to populate all necessary spectator states over competing alternatives.

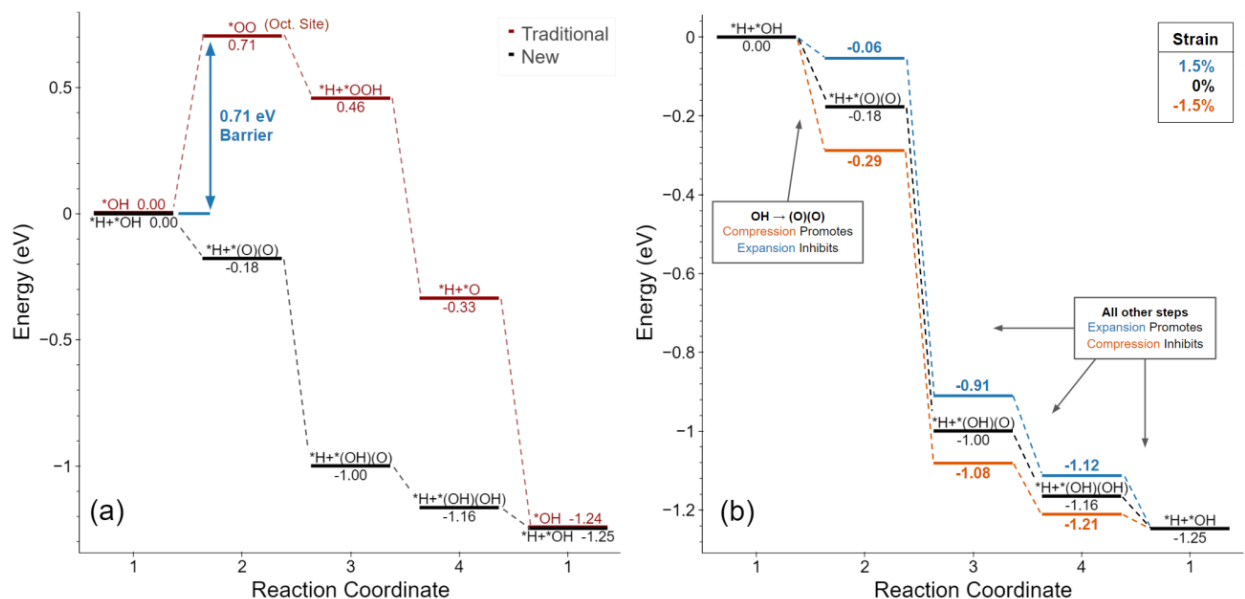

*Figure 4. Reaction intermediate energies calculated for the (100) facet at $V_{on}$ = 0.89 V vs. RHE, ordered by electron loss: (a) Values for the best possible traditional pathway (upper, red levels) and the new proposed pathway (lower, black levels), (b) Values along the proposed pathway at -1.5% (orange), 0% (black), and 1.5% (blue) tensile epitaxial strain. The reaction is cyclic, which we indicate by repeating Step 1 in the figure at an energy separation corresponding precisely to the energy we calculate to be released in one cycle of the reaction.*

## Optimality of the New Reaction Pathway

To establish that our identified pathway in Equation 3 is indeed the most energetically favorable pathway, we combine our theoretical and experimental results as follows.

First, the simplicity of State 1, the *OH state, allowed us to conduct a comprehensive search of possible configurations for this state, including those with *H and *OH as spectators. As a result, we have extremely high confidence that we identified the correct configuration and energy for State 1. Thus, we can treat the energy of this state as a fixed, well-established reference.

Moreover, we have discovered a configuration for State 4 with an energy only 0.09 eV higher than State 1 at our experimentally observed onset potential ($V_{on}$). Since the energy of State 1 is fixed, any potential lower-energy State 4 configuration must then have a drop of less than 0.09 eV when transitioning to State 1 at $V_{on}$. However, this energy drop cannot become negative at $V_{on}$, as that would inhibit the transition from State 4 to State 1, contradicting experimental results. The lack of room for improvement in energy strongly suggests that any alternative configuration for State 4 would be at most a slight modification of our proposed configuration, perhaps involving some additional, more distant and thus less relevant spectator species.

Next, we can consider State 3 using the same reasoning as with State 4. Considering the now established energy of State 4 and the small 0.16 eV energy drop from State 3 to State 4, there is likewise little room for significant improvement over our identified configuration for State 3. Therefore, any changes to our configuration for State 3 would likely be minor as well.

Finally, State 2, unlike State 3 and State 4, theoretically could have a new configuration which drops in energy by as much as 0.7 eV without necessarily inhibiting the reaction. In theory, there could exist a qualitatively different State 2 at a significantly lower energy than our proposed split *H + *(O)(O) configuration. While we cannot definitively rule out this possibility, after a comprehensive search of potential *OO configurations, we found our proposed State 2 as the most energetically favorable and kinetically feasible intermediary between State 1 and State 3. This evidence strongly suggests that we have identified the correct configuration for State 2.

However, even if an improved configuration for State 2 were discovered, our primary conclusion about the significance of split intermediate configurations for the ORR pathway still stands because States 3 and 4 are also both split intermediates. Moreover, the transitions from States 3 to 4 to 1 first become feasible at the correct, experimentally observed onset potential, and the small drops of these transitions imply they very likely are the rate-limiting steps for the reaction. Consequently, we have high confidence in our identification of the key steps in the alkaline ORR reaction pathway on $Co_3O_4$(100) surface.

## Reaction Kinetics

Figure 1 presents the proposed kinetics for our pathway as follows:

1. **\*H + \*OH→\* + \*(OH)(O):** Initially, *OH is adsorbed on the $Co^{2+}$ site and a spectator H is adsorbed on one of the four nearby surface O sites. The following motions then occur simultaneously in a coordinated fashion. An $O_2$ from the solution displaces the adsorbed OH, cracks on the $Co^{2+}$ site with one of the O's forming a hydrogen bond with the spectator H, which lifts toward that O. ( *H + *(O)(O) → * + *(OH)(O) ) Meanwhile, the departing OH takes with it an $e^-$ from the electrode to form $OH^-$.
2. **\* + \*(OH)(O)→\* + \*(OH)(OH):** An $H^+$ is taken from an $H_2O$ to form a covalent bond with the isolated adsorbed oxygen from the * + *(OH)(O) configuration, leaving an $OH^-$ behind in solution. Meanwhile, the $H^+$ gathers an $e^-$ from the electrode, forms a covalent bond with the O, and leans over to form an additional hydrogen bond with a nearby surface oxygen atom.
3. **\* + \*(OH)(OH)→\*H + \*(OH)(OH):** Either one of the adsorbed (OH) fragments takes an $H^+$ from an $H_2O$ in solution. This $H^+$ absorbs an $e^-$ from the surface to bond covalently with the corresponding OH while that OH transfers its original H to a surface oxygen site, converting its covalent bond into a hydrogen bond and producing a spectator H. Meanwhile, the newly adsorbed H from the solution forms a hydrogen bond with another nearby surface oxygen site.

4. **\*H + \*(OH)(OH) → \*H + \*OH:** Finally, one of the attached (OH) groups leaves the surface, bringing the final e$^-$ with it into solution, thereby restoring the original state in the catalytic sequence.

## Impact of epitaxial strain

By growing $Co_3O_4$ as the shell of core-shell nanoparticles, we can induce epitaxial strain and tune the energies of the reaction intermediates to improve fuel-cell performance. The change in energy per adsorbate $\Delta E^{(a)}$ with epitaxial lattice strain $\varepsilon_{sys}$ can be computed from *ab initio* calculations directly as

$\Delta E^{(a)} = \Omega \ \text{tr} \ ([\sigma_{3d}^{(a)} - \sigma_{3d}^{(0)}]_{2\times 2}[\varepsilon_{sys}]_{2\times 2})$,

where $[...]_{2\times 2}$ indicates restriction of a three-dimensional tensor to the two-dimensional space of the plane of the material surface, $\sigma_{3d}^{(0)}$ and $\sigma_{3d}^{(a)}$ are, respectively, the three-dimensional stress tensors from *ab initio* periodic supercell calculations of the clean surface and the surface with one adsorbate, and $\Omega$ is the volume of the supercell in which the three-dimensional calculations are performed. Defining the adsorbate surface-stress tensor as $s^{(a)} \equiv \Omega \ [\sigma_{3d}^{(a)} - \sigma_{3d}^{(0)}]_{2\times 2}$, we can write the change in energy per unit area of the surface due to adsorbates as $\Delta E_{ads}/\Delta A = \eta \ \text{tr} \ s^{(a)} \ [\varepsilon_{sys}]_{2\times 2}$, where $\eta$ is the density of adsorbates on the surface per unit area, and $s^{(a)}$ is thereby the surface chemical stress tensor associated with the adsorbate.

With the above surface-strain induced shifts in intermediate state energies, shifts in the onset potential ($V_{on}$) can also be expected. To analyze these shifts in $V_{on}$, we first observe that our identified pathway is sequential without branching. Consequently, the reaction will be switched on and off by whichever step $k \rightarrow k+1$ is rate-limiting at $V_{on}$. For a preliminary analysis, we next make the reasonable assumption that, for small shifts in strain, the final net drop in energy of the rate-limiting step, $E_k - E_{k+1}$, will remain constant at the reaction onset. From the previous paragraph, we have that the application of epitaxial strain $\varepsilon$ changes the energy drop between successive states to $E_k - E_{k+1} + \text{tr}(s^{(k)} - s^{(k+1)}) \cdot \varepsilon$, so that, to maintain the same final net energy drop, the onset potential must shift by $\Delta V = \text{tr}(s^{(k)} - s^{(k+1)}) \cdot \varepsilon \ / \ (ne)$, where $n$ is the properly signed number of electrons transferred during the step. Thus, we can expect a rate of change of $V_{on}$ with respect to epitaxial strain of $dV_{on}/d\varepsilon = \text{tr}(s^{(k)} - s^{(k+1)}) \ / \ (ne)$. Because each reaction step will have a characteristic value for the derivative, by comparing calculated values and experimental measurements of $dV_{on}/d\varepsilon$, one can both confirm that the identified step is present in the actual reaction pathway and determine that it is the rate limiting step. Thus, through this electrochemical strain spectroscopy (ESS), direct comparisons and identifications can be made between experiments and *ab initio* calculations.

Table 1 presents our predicted onset potential derivatives $dV_{on}/d\varepsilon$ and predicted $V_{on}$ potentials at -1.5% and 1.5% strain for the case of each reaction step being the possible rate limiting step for our proposed ORR pathway. Notably, the sign of the strain derivative for the 1→ 2 step is negative, making this the only rate limiting step which improves with compressive strain. Moreover, the magnitude of the strain derivative for the 2 → 3 step is half that of the 3 → 4 and 4 → 1 steps, making the 2 → 3 step also easy to distinguish from the others.

Finally, from the fact that we have examples of both signs for the onset potential derivative $dV_{on}/d\varepsilon$ with respect to strain, we learn that performance cannot be expected to continue to improve with increasing strain and that there will always be an optimal magnitude for the strain. This is because, regardless of the sign of strain preferred by the rate limiting step, as the rate-limiting step improves with strain, there are always other reactions steps with derivatives of opposite sign that therefore become less and less favorable as the magnitude of the strain increases, until one of these steps becomes the rate limiting step. The optimal strain will then be the point of cross-over between the original step and this new step being the actual rate limiting step.

| Intermediate | $dE_k/d\varepsilon$ (meV / %) | Rate-limiting Step | $dV_{on}/d\varepsilon$ (mV / %) | $V_{on}$ at -1.5% Strain (V) | $V_{on}$ at 1.5% Strain (V) |
|---|---|---|---|---|---|
| 1  *H + *OH | -108 | 1 → 2 | -75 | 1.00 | 0.78 |
| 2  * + *(OH)(O) | -33 | 2 → 3 | 18 | 0.86 | 0.92 |
| 3  * + *(OH)(OH) | -52 | 3 → 4 | 26 | 0.85 | 0.93 |
| 4  *H + *(OH)(OH) | -78 | 4 → 1 | 30 | 0.84 | 0.94 |

*Table 1. Predictions of impact of epitaxial strain on energies of intermediate states and on the onset potential ($V_{on}$) for each case of one of the reaction steps being the actual rate-limiting step. At present, 3 → 4 and 4 → 1, having the lowest energy drops at $V_{on}$, appear to be the rate-limiting steps.*

## Conclusion

This work characterizes the oxygen-reduction reaction (ORR) on the (100) surface of cubic $Co_3O_4$. To explain the experimentally observed onset potential, we performed solvated joint density-functional theory (JDFT) calculations using a detailed computational hydrogen electrode (CHE) method. Our results ultimately led us to propose a new spinel oxide alkaline ORR pathway, which, unlike the pathway typically assumed for metal oxides, exhibits *ab initio* free energy values in strong agreement with experimental onset potentials for $Co_3O_4$ cubic nanoparticles. The newly proposed pathway shows that PGM-like split (or "cracked") intermediate states in the form *(O)(O), *(O)(OH), and *(OH)(OH), which had not been considered previously for these systems, mediate the reaction by substantially lowering the reaction barrier though hydrogen bonding with induced spectator surface *H atoms. Using our novel electrochemical strain spectroscopy (ESS) technique, we also predict the effects of epitaxial strain on the reaction to characterize the strain signature of the limiting step. Having now identified the controlling pathway for ORR on spinel oxides and demonstrated the capability of first principles calculations to explain the onset potential, the door is now open for the *ab initio* exploration of next generation of fuel cell electrocatalysts.

## Acknowledgments

This work was primarily supported by the Center for Alkaline-Based Energy Solutions (CABES), part of the Energy Frontier Research Center (EFRC) program supported by the U.S. Department of Energy, under grant DE-SC-0019445. This work made use of TEM of the Cornell Center for Materials Research Shared Facilities which are supported through the NSF MRSEC program (DMR-1719875). We thank Malcolm (Mick) Thomas at CCMR for help with Nion Ultra STEM

This work was also supported by the Department of Defense (DoD) through the National Defense Science & Engineering Graduate (NDSEG) Fellowship Program. We thank Dr. Kyle Grew of the U.S. Army Research Laboratory for his mentorship throughout the NDSEG program.

# References

1. Stamenkovic, V. R. *et al.* Improved oxygen reduction activity on Pt3Ni(111) via increased surface site availability. *Science (1979)* **315**, 493–497 (2007).

2. Firouzjaie, H. A. & Mustain, W. E. Catalytic Advantages, Challenges, and Priorities in Alkaline Membrane Fuel Cells. *ACS Catalysis* vol. 10 225–234 Preprint at https://doi.org/10.1021/acscatal.9b03892 (2020).

3. Ge, X. *et al.* Oxygen Reduction in Alkaline Media: From Mechanisms to Recent Advances of Catalysts. *ACS Catal* **5**, 4643–4667 (2015).

4. Banham, D. & Ye, S. Current status and future development of catalyst materials and catalyst layers for proton exchange membrane fuel cells: An industrial perspective. *ACS Energy Lett* **2**, 629–638 (2017).

5. Gottesfeld, S. *et al.* Anion exchange membrane fuel cells: Current status and remaining challenges. *J Power Sources* **375**, 170–184 (2018).

6. Dekel, D. R. Review of cell performance in anion exchange membrane fuel cells. *J Power Sources* **375**, 158–169 (2018).

7. Chung, H. T. *et al.* Direct atomic-level insight into the active sites of a high-performance PGM-free ORR catalyst. *Science (1979)* **357**, 479–484 (2017).

8. Yang, Y., Zeng, R., Xiong, Y., Disalvo, F. J. & Abruña, H. D. Cobalt-Based Nitride-Core Oxide-Shell Oxygen Reduction Electrocatalysts. *J Am Chem Soc* **141**, 19241–19245 (2019).

9. Yang, Y. *et al.* In Situ X-ray Absorption Spectroscopy of a Synergistic Co-Mn Oxide Catalyst for the Oxygen Reduction Reaction. *J Am Chem Soc* **141**, 1463–1466 (2019).

10. Seo, B. *et al.* Size-Dependent Activity Trends Combined with in Situ X-ray Absorption Spectroscopy Reveal Insights into Cobalt Oxide/Carbon Nanotube-Catalyzed Bifunctional Oxygen Electrocatalysis. *ACS Catal* **6**, 4347–4355 (2016).

11. Suntivich, J. *et al.* Design principles for oxygen-reduction activity on perovskite oxide catalysts for fuel cells and metal–air batteries. *Nature Chemistry 2011 3:7* **3**, 546–550 (2011).

12. Ramaswamy, N., Tylus, U., Jia, Q. & Mukerjee, S. Activity descriptor identification for oxygen reduction on nonprecious electrocatalysts: Linking surface science to coordination chemistry. *J Am Chem Soc* **135**, 15443–15449 (2013).

13. Wei, C. *et al.* Cations in Octahedral Sites: A Descriptor for Oxygen Electrocatalysis on Transition-Metal Spinels. *Advanced Materials* **29**, 1606800 (2017).

14. Li, C. *et al.* Phase and composition controllable synthesis of cobalt manganese spinel nanoparticles towards efficient oxygen electrocatalysis. *Nature Communications 2015 6:1* **6**, 1–8 (2015).

15. Duan, J., Chen, S., Dai, S. & Qiao, S. Z. Shape Control of Mn3O4 Nanoparticles on Nitrogen-Doped Graphene for Enhanced Oxygen Reduction Activity. *Adv Funct Mater* **24**, 2072–2078 (2014).

16. Gao, R. *et al.* Facet-dependent electrocatalytic performance of Co3O4 for rechargeable Li-O2 battery. *Journal of Physical Chemistry C* **119**, 4516–4523 (2015).

17. Su, D., Dou, S. & Wang, G. Single Crystalline Co3O4 Nanocrystals Exposed with Different Crystal Planes for Li-O2 Batteries. *Scientific Reports 2014 4:1* **4**, 1–9 (2014).

18. Rong, X. & Kolpak, A. M. Ab initio approach for prediction of oxide surface structure, stoichiometry, and electrocatalytic activity in aqueous solution. *Journal of Physical Chemistry Letters* **6**, 1785–1789 (2015).

19. Wang, Y. & Cheng, H. P. Oxygen reduction activity on perovskite oxide surfaces: A comparative first-principles study of LaMnO3, LaFeO3, and LaCrO 3. *Journal of Physical Chemistry C* **117**, 2106–2112 (2013).

20. Goodenough, J. B. & Cushing, B. L. Oxide-based ORR catalysts. *Handbook of Fuel Cells* (2010) doi:10.1002/9780470974001.F205040.

21. Sundararaman, R. *et al.* JDFTx: Software for joint density-functional theory. *SoftwareX* **6**, 278–284 (2017).

22. Gunceler, D., Letchworth-Weaver, K., Sundararaman, R., Schwarz, K. A. & Arias, T. A. The importance of nonlinear fluid response in joint density-functional theory studies of battery systems. *Model Simul Mat Sci Eng* **21**, 074005 (2013).

23. Yang, Y. *et al.* Octahedral spinel electrocatalysts for alkaline fuel cells. *Proc Natl Acad Sci U S A* **116**, 24425–24432 (2019).

24. Zasada, F., Piskorz, W. & Sojka, Z. Cobalt Spinel at Various Redox Conditions: DFT+U Investigations into the Structure and Surface Thermodynamics of the (100) Facet. *Journal of Physical Chemistry C* **119**, 19180–19191 (2015).

25. Agmon, N. *The Grotthuss mechanism*. *Chemical Physics Letters* vol. 244 (1995).

26. Momma, K. & Izumi, F. VESTA 3 for three-dimensional visualization of crystal, volumetric and morphology data. *J Appl Crystallogr* **44**, 1272–1276 (2011).

27. Perdew, J. P., Burke, K. & Ernzerhof, M. *Generalized Gradient Approximation Made Simple*. (1996).

28. Garrity, K. F., Bennett, J. W., Rabe, K. M. & Vanderbilt, D. Pseudopotentials for high-throughput DFT calculations. *Comput Mater Sci* **81**, 446–452 (2014).

29. Sundararaman, R., Gunceler, D. & Arias, T. A. Weighted-density functionals for cavity formation and dispersion energies in continuum solvation models. *J Chem Phys* **141**, 134105 (2014).

30. Nørskov, J. K. *et al.* Origin of the overpotential for oxygen reduction at a fuel-cell cathode. *Journal of Physical Chemistry B* **108**, 17886–17892 (2004).

31. Zasada, F., Gryboś, J., Piskorz, W. & Sojka, Z. Cobalt Spinel (111) Facets of Various Stoichiometry - DFT+U and Ab Initio Thermodynamic Investigations. *Journal of Physical Chemistry C* **122**, 2866–2879 (2018).

32. Groß, A. Reversible vs Standard Hydrogen Electrode Scale in Interfacial Electrochemistry from a Theoretician's Atomistic Point of View. *Journal of Physical Chemistry C* **126**, 11439–11446 (2022).

33. Jerkiewicz, G. Standard and Reversible Hydrogen Electrodes: Theory, Design, Operation, and Applications. *ACS Catal* **10**, 8409–8417 (2020).

34. Cai, Y. & Anderson, A. B. The reversible hydrogen electrode: Potential-dependent activation energies over platinum from quantum theory. *Journal of Physical Chemistry B* **108**, 9829–9833 (2004).

35. Oberhofer, H. & Reuter, K. First-principles thermodynamic screening approach to photo-catalytic water splitting with co-catalysts. *J Chem Phys* **139**, 044710 (2013).

36. Valdés, Á., Qu, Z. W., Kroes, G. J., Rossmeisl, J. & Nørskov, J. K. Oxidation and photo-oxidation of water on TiO2 surface. *Journal of Physical Chemistry C* **112**, 9872–9879 (2008).

37. Gamache, R. R. *et al.* Total internal partition sums for the HITRAN2020 database. *J Quant Spectrosc Radiat Transf* **271**, (2021).

38. Capitelli, M., European Space Agency. & European Space Research and Technology Centre. *Tables of internal partition functions and thermodynamic properties of high-temperature Mars-atmosphere species from 50K to 50,000K*. (ESA Publications Division, ESTEC, 2005).

# Supplemental Information

## S1: *Ab Initio* Calculation Methods

We perform all calculations using joint density-functional theory (JDFT) in the JDFTx plane-wave implementation,[21] approximating the exchange-correlation energy using the GGA-PBE functional.[27] We use GBRV ultrasoft pseudopotentials (USPP),[28] which allow us to employ a plane-wave cutoff of 20 Hartree (~544 eV) and a charge-density cutoff of 100 Hartree (~2,720 eV). To capture the effects of the solvent, including the dielectric and ionic response, we employ the GLSSA13 non-linear Polarizable Continuum Model (PCM),[22,29] and we iteratively converge all *ab initio* calculations with respect to electronic and ionic coordinates to within ~1 mH (0.027 eV) for each configuration and control k-point and slab-thickness convergences to within 0.1 eV. Finally, to compute free-energy differences between reaction steps, we use the computational hydrogen electrode (CHE)[30] in the manner discussed in S2, taking care to include rotational contributions to the molecular free energies.

For calculation of reaction intermediate states on the surface, we use the supercell slab method with slabs constructed from conventional cubic unit cells of lattice constant 8.084 Å.[24,31] For the (100) facet, we used a stoichiometrically terminated 1 × 1 slab of thickness 10.105 Å with vacuum spacing of 13.229 Å between slabs, for a final supercell dimension of 8.084 Å × 8.084 Å × 23.334 Å. We use a Brillouin-zone sampling mesh of 2x2x1 k-points. To confirm convergence with k-point sampling, we examined all key reaction intermediate configurations (defined as those within 0.2 eV of the minimum energy configuration found for each reaction step as identified below) at an increased Brillouin zone sampling of 3x3x1 k-points. We additionally evaluated all key reaction intermediate configurations at the 0.5 eV threshold at *decreased* slab thickness of 7.55 Å. All key configurations changed by less than 0.1 eV with these variations in parameters. Therefore, we expect our *ab initio* calculations to be converged to within 0.1 eV.

## S2: Reaction Energy Calculations

Direct comparison of reaction energies between different steps of the reaction requires knowledge of the energies of the reactants consumed and the products created between steps. To account for such changes, we add the product $N_s \mu_s$ of the number $N_s$ and chemical potential $\mu_s$ of each product/reactant species *s* remaining in the solution to the computed energy of each surface configuration. We calculate the needed chemical potentials using a refinement of the computational hydrogen electrode (CHE) method to represent the experimental reversible hydrogen electrode (RHE).[30,32–36] In the present manuscript, we evaluate the needed chemical potentials by including the electronic, solvation, configurational, and rotational free energy contributions to the chemical potentials of all reactants and products. When using the CHE, we include the rotational parts of the internal free energy because of their contributions at the 0.1 eV level, as seen in Table S1. However, we neglect vibrational and zero-point energy contributions at room temperature, which we estimate to be a correction on the order of 0.01 eV or less.[37,38]

| Species | Phase | Total DFT Free Energy $E_{DFT}$ (eV) | Configurational $\Delta G_{conf}$ (eV) | Rotational $\Delta G_{rot}$ (eV) |
|---|---|---|---|---|
| $O_2$ | Gas | -871.402 | -0.486 | -0.111 |
| $H_2$ | Gas | -31.705 | -0.379 | -0.017 |
| $H_2O$ | Liquid | -470.218 | -0.278 | -0.097 |

Table S1. Free-energy contributions of reactants and products. Note that these DFT free energies include the energies of the pseudopotential cores that are needed for proper energy cancellation.

For the reactions considered here, we require knowledge of the following chemical potentials: $\mu(e^-)$, $\mu(O_2(g))$, $\mu(H_2O(\ell))$, $\mu(OH^-)$. We determine these as follows. **$\mu(H_2O(\ell))$:** To best benefit from cancellation of errors between calculations of different configurations, we take $\mu(H_2O(\ell))$ to be the total energy of an isolated, *solvated* water molecule as calculated within our DFT framework plus the relevant configurational and rotational free energies for water at 25 C. **$\mu(O_2(g))$:** We compute $\mu(O_2(g))$ to be the total DFT energy of an isolated $O_2$ molecule in vacuum plus the relevant configurational and rotational free energies at 25

C. **μ(e⁻):** By definition, the potential of the RHE is set to zero, so that $\mu(e^-_{RHE}) = 0$ at the RHE. Correspondingly, at the working electrode, $\mu(e^-) = -eV$, where $V$ is the voltage between the working and hydrogen electrodes, and $e$ is the fundamental charge. **μ(OH⁻):** To determine the remaining chemical potential $\mu(OH^-)$, we employ the local water equilibrium, $OH^- + H^+ \leftrightarrows H_2O(\ell)$, so that we take $\mu(OH^-) = \mu(H_2O(\ell)) - \mu(H^+)$, with $\mu(H_2O(\ell))$ determined above. In order to determine $\mu(H^+)$, we note that, by definition, $H_2(g) \leftrightarrows 2H^+ + 2e^-_{RHE}$ at the RHE when held under the same pH conditions as the working electrode. Thus $\mu(H_2(g)) = 2\cdot\mu(H^+) + 2\cdot\mu(e^-_{RHE}) = 2\cdot\mu(H^+)$, allowing us to relate $\mu(H^+) = \frac{1}{2}\mu(H_2(g))$. Finally, we take $\mu(H_2(g))$ to be our DFT energy of an isolated $H_2$ molecule in vacuum plus the relevant configurational and rotational free energies at 25 C. Equation (S1) summarizes the determination of the chemical potentials employed to produce the results reported below.

$\mu(e^-) = -eV$

$\mu(O_2(g)) = E_{DFT}(O_2) + \Delta G_{conf}(O_2(g)) + \Delta G_{rot}(O_2(g))$

$\mu(H_2O(\ell)) = E^{solv}_{DFT}(H_2O) + \Delta G_{conf}(H_2O(\ell)) + \Delta G_{rot}(H_2O(\ell))$

$\mu(OH^-) = \mu(H_2O(\ell)) - \frac{1}{2}[E_{DFT}(H_2) + \Delta G_{conf}(H_2(g)) + \Delta G_{rot}(H_2(g))]$

(S1)

## S3: Synthesis and Electrochemical Measurements

We used the hydrothermal synthesis method to synthesize the $Co_3O_4$ NPs. First $Co(NO_3)_2 \cdot 6H_2O$ and NaOH were dissolved in water. The solution was then transferred to a Teflon lined autoclave reactor and heated at 180°C for 5 hours. The resulting product was collected by centrifugation and washed 2 times with ethanol and water. The product was then supported on carbon powder (Vulcan-72R). We used XRD with a Cu $K_\alpha$ X-ray source and a Rigaku Ultima IV diffractometer to determine the crystal structure of the synthesized NPs. We conducted bright field STEM imaging using a fifth-order aberration-corrected STEM (Nion-UltraSTEM-100) operating at 100 KeV.

We then performed electrochemical measurements in 1 M KOH on a potentiostat. We dispersed prepared $Co_3O_4$/C catalyst in 1 ml of 0.05 wt% Nafion/ethanol and sonicated for 10 minutes before drop casting on a RRDE with a glassy carbon disk and a platinum ring. We used a carbon rod and a saturated Ag/AgCl electrode as a counter and a reference electrode, respectively. Finally, we conducted the ORR measurement in oxygen saturated electrolyte at 5 mVs$^{-1}$.